\documentclass{article}

\usepackage{arxiv}

\usepackage[utf8]{inputenc} 
\usepackage[T1]{fontenc}    
\usepackage{hyperref}       
\usepackage{url}            
\usepackage{booktabs}       
\usepackage{amsfonts}       
\usepackage{nicefrac}       
\usepackage{microtype}      
\usepackage{lipsum}
\usepackage{graphicx}
\graphicspath{ {./images/} }
\usepackage{booktabs}
\usepackage{tabularx}
\usepackage{colortbl}
\usepackage{natbib}

\title{Self-Regulated Learning in Essay Writing: Consistency of Strategies and Impact on Outcomes}

\author{
 Gloria Fernández-Nieto \\
  Department of Data Science \& AI\\
  Monash University\\
  Australia, Victoria\\
  \texttt{gloriamilena.fernandeznieto@monash.edu} \\
   \And
Kiyoshige Garcés \\
    Department of Data Science \& AI\\
  Monash University\\
  Australia, Victoria\\
  \texttt{Oscar.GarcesAparicio@monash.edu} \\
  \And
 Mladen Raković \\
  Department of Data Science \& AI\\
  Monash University\\
  Australia, Victoria\\
  \texttt{mladen.rakovic@monash.edu} \\
  \And
 Tongguang Li \\
  Department of Data Science \& AI\\
  Monash University \\
  Australia, Victoria\\
  \texttt{tongguang.li@monash.edu} \\
 \And
 Xinyu Li \\
  Department of Data Science \& AI\\
  Monash University \\
  Australia, Victoria\\
  \texttt{Xinyu.Li1@monash.edu} \\
\And
 Linxuan Zhao\\
  Department of Data Science \& AI\\
  Monash University \\
  Australia, Victoria\\
  \texttt{linxuan.zhao@monash.edu} \\
  \And
 Dragan Gašević\\
  Department of Data Science \& AI\\
  Monash University \\
  Australia, Victoria\\
  \texttt{dragan.gasevic@monash.edu} \\
}

\begin{document}
\maketitle
\begin{abstract}
\textbf{Background:} Abilities for effective self-regulated learning (SRL) are critical for lifelong learning, particularly during adolescence when these skills consolidate and strongly influence future learning. Their importance has grown with the rise of online and blended education. Yet, little is known about how secondary school students self-regulate in online environments, how their SRL processes and strategies evolve, or how they affect outcomes. In secondary education, understanding these processes can reveal patterns and indicators of learning success, informing the design of online support mechanisms. Evidence from repeated-measures designs remains scarce.
\textbf{Objectives:} This study aims to examine how secondary school students enact SRL strategies during online essay writing, how these strategies change over time, and how they relate to learning outcomes.
\textbf{Methods: } We analysed metacognition-related trace data collected from secondary students during a two-wave online essay-writing task conducted one week apart in two Colombian 
schools ($N = 93$ for session 1, $N = 95$ for session 2) via a digital learning platform. Using a combination of process mining and unsupervised machine learning techniques, we identified dominant SRL strategies grounded in established SRL processes and examined their stability and association with learning outcomes.
\textbf{Results and conclusions} Three dominant SRL strategies were identified. Results showed variability: many students remained in or shifted to \textit{Read first, write next}, while none used \textit{Write intensively, read selectively} in session 2. Although less common, latter strategy was positively associated with learning outcomes.
\end{abstract}


\section{Introduction}

Being a proficient learner requires a set of skills that enable the adaptation of learning processes to different contexts \cite{Zimmerman01011990, winne1998studying}. This set of skills is commonly referred to as self-regulated learning (SRL), which is defined as a cyclical, constructive, and active process encompassing cognitive, metacognitive, motivational, emotional, and behavioural dimensions~\cite{schunk2011handbook}. These skills are considered essential in contemporary education and society, as they provide the foundation for lifelong learning~\cite{fan2022improving}. As a result, the Learning Analytics (LA) community has increasingly focused on research aimed at understanding SRL in educational contexts~\cite{fan2021learning}.

With the current proliferation of online and blended teaching, supporting SRL processes in digital environments has become increasingly important for academic success~\cite{broadbent2015self}. Understanding students' SRL strategies -- i.e., sequence of learning processes that students enact and coordinate to accomplish their learning goals\cite{winne1998studying,srivastava2022effects} -- can provide valuable insights into learning and inform the provision of targeted support in online contexts. To this end, trace data collected from online and blended learning environments have become instrumental in capturing students' SRL processes and strategies~\cite{molenaar2023measuring}. Researchers have developed several frameworks to aid understanding of SRL processes using trace data, including the frameworks that identify processes at both macro and micro levels (e.g., \cite{fan2022towards, azevedo2008externally}). Such understanding of SRL can inform the development of support mechanisms (e.g., adaptive learning systems or SRL coaching) that help students with tasks such as monitoring or planning their learning processes \cite{lim2024students}. Nevertheless, many students continue to struggle with self-regulation in online settings, often due to a limited understanding of their own SRL strategies and how these strategies influence learning outcomes~\cite{azevedo2008externally, heikkinen2023supporting, li2024analytics}. Gaining insights into how students enact SRL strategies in different task is critical, as it can inform researchers and educators to design more effective learning supports to help students succeed in online environments~\cite{Halima2025}. Given that SRL is a set of learning processes that dynamically unfold during a learning session and that are specific to each individual student, researchers have been increasingly emphasising the importance of detecting SRL strategies as individual students' sequences of SRL processes~\cite{srivastava2022effects,osakweReinforcementLearningAutomatic2023, Garces2025}. For example, empirical studies have examined these sequences to identify optimal orders of SRL processes linked to learning outcomes \cite{osakweReinforcementLearningAutomatic2023}. However, most of this work has been conducted in higher education, with far less attention to secondary school contexts, even though SRL is recognised as a critical aspect of academic success and personal development in secondary schools~\cite{dignath2008components, kesuma2021analysis}. Findings from  Weil et al. \cite{Weil2013} suggest that adolescence is a critical period for the development of metacognitive skills that are central to productive SRL \cite{winne1998studying}. Specifically, metacognitive skills peak in late adolescence and remain stable into adulthood. Therefore, understanding how SRL processes are enacted in secondary education is crucial to ensure appropriate support throughout students' educational journeys and to foster the proper development of metacognitive skills~\cite{cheng2025self}.

To ensure appropriate support in students' educational journeys, it is essential to understand how students adapt SRL strategies over time, aligning with the adaptation phase of SRL proposed by Winne and
Hadwin \cite{winne1998studying}. To investigate such adaptations, LA approaches based on process mining can be applied to systematically compare repeated measurements and analyse the SRL strategies enacted by students across study sessions -- e.g., whether strategies used in one session are sustained or adapted in the next~\cite{winne1998studying, Higgins2023}. 
Although understanding SRL strategies across repeated measurements (e.g., micro-longitudinal studies) is highly relevant, such studies remain notably scarce \cite{Higgins2023}. To date, most comparisons across measures have focused on higher education \cite{Stanton2019}, relying on repetitive surveys \cite{Esnaashari2023} or think-aloud protocols \cite{van2010development}. Some seminal work harnessed trace data to provide important insights into students' SRL processes in both higher and secondary education. For instance, research has shown that learning outcomes in higher education are related to the SRL strategies that student enact and adapt over time \cite{Saqr2023}. Other studies have demonstrated that metacognitive skills make a significant contribution to learning outcomes \cite{van2014metacognitive} and that the development of metacognitive skills is specific to individual students \cite{van2010development}. Moreover, students have been found that students to diversify their SRL strategies over the course of a semester \cite{matcha2019analytics,rakovic2023network}. Extending this perspective, recent work has shown that both secondary and university students demonstrate diverse SRL processes during writing activities~\cite{cheng2025self}. Yet, there is limited understanding of (1) how SRL strategies develop over time, (2) how consistently students apply SRL strategies in different tasks, (3) how SRL strategies relate to students' learning outcomes, and (4) how such an understanding can guide the provision of online SRL support \cite{cheng2025self, Halima2025}.

The current study was conducted to addresses the above gaps by analysing SRL strategies among secondary school students in [anonymised country] 
engaged in multi-source essay-writing tasks on a custom-built digital learning platform. Using process mining techniques, we examined data collected across two sessions conducted one week apart ($N = 93$ in session 1 and $N = 95$ in session 2). Specifically, the study investigated (i) the SRL strategies students employ, (ii) the extent to which these strategies remain consistent across tasks, and (iii) how they relate to learning outcomes. The findings revealed variability in strategy use across sessions, with a preference for a \textit{Read first, write next} strategy, and highlighted how less common strategies, such as \textit{Write Intensively, Read Selectively}, can be positively associated with learning outcomes. These findings contribute to a deeper understanding of secondary students’ SRL and can inform the design of effective online support mechanisms and teacher education programs.

\section{Background}


\subsection{Understanding and Measuring Self-Regulated Learning through Trace Data}

SRL is a multi-dimensional construct encompassing cognitive, metacognitive, motivational, and affective dimensions of human learning, through which learners engage in learning tasks \cite{Zimmerman01011990,panadero2017review,winne1998studying}. Researchers have proposed theoretical and methodological frameworks to understand SRL processes, from macro-level approaches \cite{fan2022towards} that focus on broad phases of SRL, to micro-level \cite{azevedo2008externally} frameworks that examine moment-to-moment interactions. Recent literature reviews \cite{deMooij2025, MOLENAAR2023} have examined what types of trace data are typically collected and how they are used to study SRL.
For example,  Fan et al. \cite{fan2022towards} built on Bannert et al.’s \cite{bannert2014process} 
theoretical SRL framework \cite{bannert2014process} to developed an iterative approach that combines theory-driven and data-driven methods to identify SRL processes.
Fan et al. \cite{fan2022towards} translated students’ digital trace data (e.g., clicks, mouse movements, and keystrokes) into meaningful learning actions (or student interactions such as read) and used a \textit{process library} to interpret these actions as SRL processes a set of rules that link observed learning actions to underlying SRL processes.
The process library includes three main categories as originally posited by Bannert et al. \cite{bannert2014process}: Metacognitive (MC), Low Cognitive (LC), and High Cognitive (HC), with SRL processes associated with each category. MC processes include orientation (O), planning (P), monitoring (M) and evaluation (E); LC processes include first reading and re-reading (R); and HC processes include elaboration/organisation (E/O). These processes extracted from trace data to 
capture intermediate steps in learning and offer valuable information about learners’ strategies.

Motivated by the need to understand students' SRL strategies, researchers have increasingly analysed digital traces of interaction with digital learning environments through the lens of SRL processes. In this paper, we use the term SRL strategies to refer to sequences of SRL processes that students enact and coordinate to achieve their goals \cite{srivastava2022effects}. 
Building on this perspective, prior studies such as  Fan et al. \cite{fan2021learning} and Matcha et al. \cite{matcha2019analytics} have proposed methods for detecting strategies from trace data by separating the process into two stages. The first stage involves identifying learning tactics, which are the specific steps or routines students use to complete a task (e.g., annotating) \cite{winne2013learning}. The second stage then maps these tactics onto broader strategies, allowing researchers to capture how students coordinate their actions into more complex patterns of self-regulated learning. In these studies, the detection of strategies relied on log data about learning actions, such as clicks or navigational steps. These are the most basic form of log data and, while informative, their interpretation is often constrained by the specific task and the platform design \cite{fan2021learning, matcha2019analytics}. Likewise, using a similar approach but grounded in Bannert et al.’s \cite{bannert2014process} theoretical SRL framework  Bannert et al. \cite{bannert2014process}, Srivastava et al. \cite{srivastava2022effects} employed higher-order SRL processes rather than log data about learning actions to map sequences of learning and explore the viability of using these processes to detect SRL strategies. In their work, Srivastava et al. \cite{srivastava2022effects} identified three theoretically meaningful SRL strategies that learners enacted while working on the essay-writing task. The three groups of SRL strategies are: \textit{Read first, write next}, \textit{Read and write simultaneously}, and \textit{Write intensively, read selectively}. 

Most empirical work on SRL processes and strategy detection has been conducted in higher education \cite{Rakovic2022}, leaving limited understanding of how these processes unfold in secondary schools. Recently, Cheng et al. \cite{cheng2025self} compared how secondary and higher education students enacted SRL processes, but not strategies. Alnashiri
et al. \cite{Halima2025} investigated SRL metacognitive processes in secondary schools with different language backgrounds, finding limited use of monitoring and control in multi-source writing; however, this study did not connect SRL processes to strategies. Similarly, Lämsä et al. \cite{Lamsa2025} analysed sequences of learning actions to detect emerging SRL processes and strategies, relying on automated detection rather than the theorised process library. Their findings aligned with SRL processes \cite{fan2022towards} and strategies \cite{srivastava2022effects} previously observed in higher education. Collectively, these studies suggest that while secondary students’ SRL processes and, to some extent, strategies can be observed, the relationship between strategy use and effective regulation remains under explored. This gap is particularly important because secondary students often struggle with SRL due to a lack of effective strategies for planning, monitoring, and evaluating their writing independently \cite{ejihpe12080069,turkben2021effect,Weil2013}. A deeper understanding of SRL processes and strategies in secondary education is therefore needed to inform the design of mechanisms that better support students’ regulation. 

To directly address the limited knowledge about SRL strategies in secondary education, in the current study, we extended the approach of Srivastava et al. \cite{srivastava2022effects} to the secondary school context, mapping sequences of learning actions into higher-order SRL processes during a controlled 
essay-writing task. Unlike prior work that examined SRL strategies across extended timeframes (e.g., entire courses), the focus of the current study was on a short, two-hour task, more representative of secondary school learning activities. Using data from two secondary schools across two sessions, we aimed to answer the following research question: \textbf{RQ1:} What self-regulated processes and strategies do secondary school students follow when working on multi-source essay-writing tasks?

%
%

\subsection{Consistency of SRL strategies in longitudinal studies}

Comparative analysis across multiple measures are essential to uncover how learners regulate their own learning and how these processes evolve over time. Such analysis help identify patterns in strategy use that may signal learning success, such as improvements in academic performance \cite{Higgins2023}. Importantly, it is not only relevant to examine which strategies students adopt, but also how consistently they apply them across tasks or instructional contexts \cite{Higgins2023}.

By examining 
patterns on activity over repeated measures, researchers can identify how students reflect on  
their SRL practices and their ability to appraise the effectiveness of their study plans \cite{srivastava2022effects, matcha2019analytics}. 
In higher education, longitudinal approaches have provided valuable insights. For example, using surveys, Stanton et al. \cite{Stanton2019} showed that third-year science students were more aware of the benefits of SRL strategies compared to first-year students. In another study, Esnaashari et al. \cite{Esnaashari2023} used repeated surveys within a semester to track students' SRL strategies, showing that learners could shift between minimal, average, and highly self-regulated strategies as they received feedback.
Similarly, studies using trace data have shown that students may diversify their strategies over the course of a semester \cite{matcha2019analytics}, balance cognitive and metacognitive processes \cite{uzir2020analytics}, and adjust their approaches in response to course design \cite{fan2021learning}. These findings highlight that SRL strategies are dynamic and context-dependent.

However, in secondary education, less is known about how SRL develops and how it relates to academic performance as students advance through school \cite{Higgins2023}. Some longitudinal studies have begun to address this gap. For instance, \citet{van2010development} used a two-year think-aloud study in mathematics and history, finding that metacognitive skills contribute to performance. Later, Van Der Stel
and Veenman \cite{van2014metacognitive} showed that metacognitive development between ages $12$ to $15$ does not progress linearly. More recently,  Cheng et al. \cite{cheng2025self} used trace data to compare SRL processes in secondary and university students; however, because the comparison involved different groups, the study did not provide insights into how consistently individual students demonstrated specific SRL processes. Although such work provides initial insights into secondary students' SRL practices, systematic longitudinal studies in secondary schools remain scarce, and little is known about the consistency of SRL strategies across tasks and over time.

Understanding whether students' SRL strategy use is stable or variable, across weeks or longer periods, provides critical insights into how SRL develops during adolescence. This represents an important gap, as secondary students are at a formative stage of developing SRL skills \cite{cheng2025self}, and their patterns of SRL may differ substantially from those of university students. To address this gap, the current study aimed to address the second research question: \textbf{RQ2:} How consistently do secondary students apply SRL strategies while working on 
essay-writing tasks?

\subsection{Association between SRL strategies and learning outcomes}

Using the SRL strategies identified in our study, we investigated their relationship with students' learning outcomes. Previous research has shown that SRL strategies can distinguish high- from low-performing students (cf., \cite{broadbent2015self}). For instance,  Saint et al. \cite{saint2020combining} observed that high-performing students tended to engage in planning and content-access strategies to enhance learning gains. Similarly, Fincham et al. \cite{fincham2018study} found that high-performing students combined content-access with metacognitive SRL strategies. Supporting these findings, Fan et al.  \cite{fan2021learning} reported that high-performing students frequently integrated content-access, evaluation, and assessment strategies throughout the semester. In contrast, Uzir et al. \cite{uzir2020analytics} suggested that students employing a diverse set of strategies, balancing preparation and assessment, outperformed those who focused on a single approach.  However, these studies focus on higher education, leaving a gap in understanding how SRL processes and strategies in secondary schools relate to learning outcomes.

While limited research has explored performance associations in secondary contexts, most of it has focused on SRL processes rather than strategies. For example, Cheng et al. \cite{cheng2025self} examined SRL processes and their association with performance among higher education and secondary school students, finding that higher education students exhibited more diverse SRL processes, whereas high-performing secondary school students relied more heavily on re-reading strategies. Similarly, Alnashiri et al. \cite{Halima2025} investigated SRL processes among secondary school students and their relationship with learning outcomes, showing that these students made limited use of metacognitive processes such as monitoring and control, and demonstrated less process diversity compared to their higher education counterparts. Extending this line of work, Garces et al. \cite{Garces2025} analysed SRL process sequences in secondary school writing tasks using Epistemic Network Analysis (ENA) and found that students with high learning outcomes (essay scores and pre- and post-tests) exhibited stronger links between elaboration/organisation and reading processes, with fewer transitions between reading and metacognitive monitoring. While these studies highlight the link between SRL processes and performance, our study examines how SRL strategies, grounded in these processes, relate to learning outcomes.

The relationship between SRL strategies and learning outcomes has been a recurring focus in higher education research. For example, Srivastava et al. \cite{srivastava2022effects} investigated the association between SRL strategies and learning outcomes among higher education students and found that those who followed a \textit{Read-first, write-next} strategy slightly outperformed peers who used other approaches. To our knowledge, no prior work has explicitly investigated this relationship in the context of secondary education. We argue that examining how SRL strategies relate to learning outcomes in secondary schools is crucial across domains and tasks, particularly in common academic activities such as writing from multiple sources. Analysing these SRL strategies may help reveal which SRL processes students should engage in to enhance both writing performance and domain-specific knowledge acquisition, while also raising important questions about whether such processes consistently yield positive outcomes across contexts. Understanding this relationship can inform future educational interventions aimed at promoting productive SRL strategies aligned with theoretically grounded SRL processes, regardless of the specific learning actions through which those processes are manifested. To address this gap, we aimed to answer our third research question:
\textbf{RQ3:} To what extent do students’ SRL strategies, grounded in SRL processes, influence their learning outcomes during multi-source essay-writing tasks?

\section{Methods}
\subsection{Study}
We conducted a two-session micro-longitudinal study in two secondary schools in Colombia 
Data were collected from the same participants across the two study sessions held one week apart. 

\subsubsection{Participants}
Session 1 involved $93$ and session 2 involved $95$ students between ages 12-15 (mean age $=14.18$ years, $SD=1.48$), 
drawn from two secondary schools in Colombia 
With ethics approval from Monash University 
Human Research Ethics Committee, all students obtained parental consent and also provided their own assent for their data to be collected and analysed in this research. 

\subsubsection{Learning environment} 
\begin{figure*}
    \centering
    \includegraphics[width=1\linewidth]{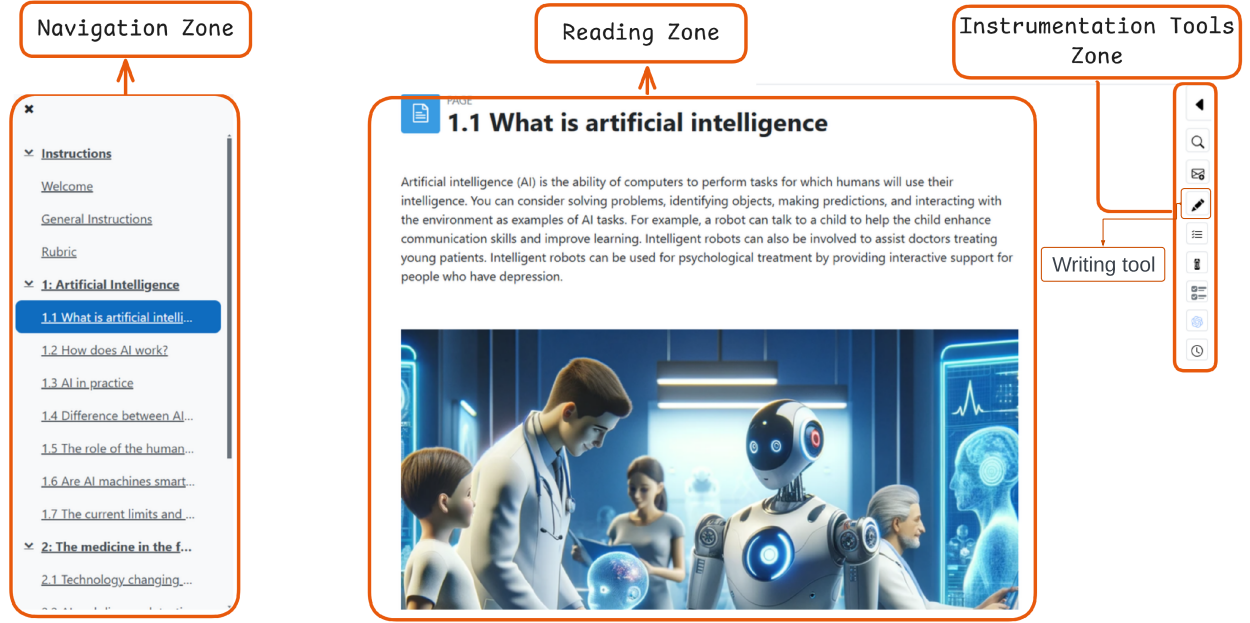}
    \caption{Screenshot of the multi-source writing task and the learning environment.}
    \label{fig:learningPlatform}
\end{figure*}

Participants used a Moodle-based learning environment to complete a writing task (see Fig. \ref{fig:learningPlatform}. The platform included reading materials accessible via a navigation bar, with texts on \textit{AI in medicine} in session 1 and \textit{AI in education} in session 2. Each session also provided task instructions and a marking rubric to familiarise students with requirements and assessment criteria. Embedded tools supported task completion, including annotation, search, timer, planning, and essay writing features.

\subsubsection{Task design}
The task had four parts. The \textit{first part} involved a pre-test to assess students' prior domain knowledge ($15$ minutes to answer $15$ multiple questions in total). The \textit{second part} provided training on the learning platform and tools ($5$ minutes). The \textit{third part} was the main task, where the students were asked to write a $300-400$ word essay in Spanish  within a $45$-minute time limit 
, using the provided tools and reading materials provided for each session. 
The design encouraged self-regulation, requiring students to identify relevant content, plan their time, and meet rubric standards rather than read everything. The \textit{fourth part} was a post-test assessing learning gains 
with the same questions as the pre-test ($15$ minutes).


\subsection{Data collection}

\subsubsection{Essay scores}

The essays produced in the \textit{third part} of the task were manually scored (\textit{ES}) by two human markers as per the following criteria in the scoring rubric: explanation of the AI concept and  examples in the context of medicine for session 1 and education for session 2 (9 points), current and future uses of AI (6 points), and word count (1 point). Scoring was based on the specific content of each session, with a maximum of 15 points per essay. To establish scoring reliability, two researchers, both native Spanish speakers, independently scored $10\%$ of the essays, aiming for a Cohen’s \textit{kappa} of $0.8$ for each criterion. After three rounds of sampling, all criteria met the threshold. The remaining essays were then divided between the two researchers for independent scoring. 

\subsubsection{Pre-test and Post-test scores}
The \textit{first part} of the task assessed students’ prior knowledge, and the \textit{fourth part} measured learning gains. In session 1, the tests had a maximum score of $15$, while in session 2 the maximum was $10$. To enable comparison of prior and post knowledge across sessions, we normalised the outcomes within each session.

\subsubsection{Mapping from raw data to SRL processes}

\begin{figure*}
    \centering
    \includegraphics[width=1\linewidth]{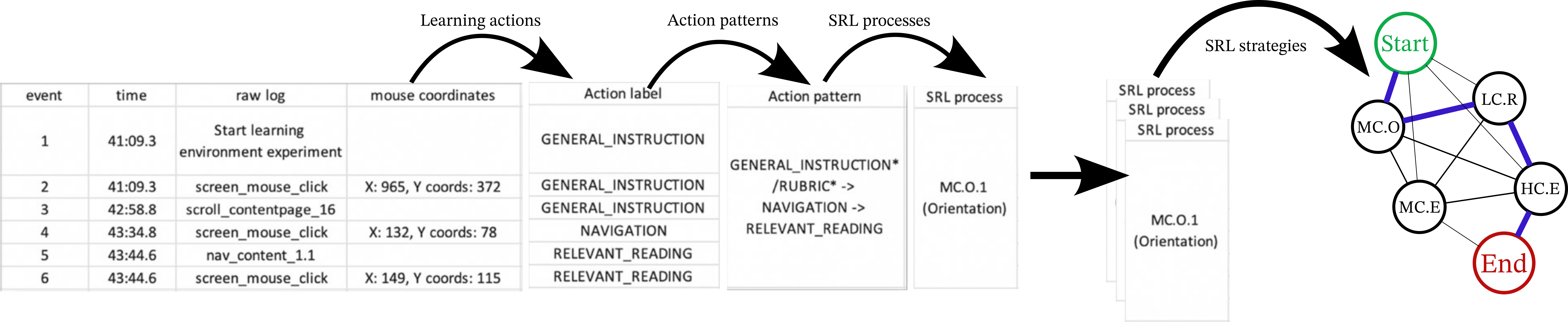}
    \caption{Process from students' trace data to SRL strategies.}
    \label{fig:srlStrategiesProcessesMapper}
\end{figure*}

The raw log data collected from the online tool included three time-stamped components: (1) navigation logs, (2) mouse traces (clicks, movements, and scrolls), and (3) keyboard strokes, all captured during the \textit{third part} of each session. We used an SRL process trace parser, informed by the framework of Bannert et al. \cite{bannert2014process}, which consisted of (i) an action library that converted raw log data into learning actions, and (ii) a process library that mapped sequential patterns of learning actions onto SRL processes (Figure \ref{fig:srlStrategiesProcessesMapper}). The validity of this parser for detecting SRL processes has been established in prior work using think-aloud and eye-tracking data \cite{fan2022towards, fan2022improving}.


Based on the theoretical framework proposed by Bannert et al. \cite{bannert2014process}, we characterised SRL in the 
writing task with three major categories (metacognition, low\_cognition, and high\_cognition) and defined seven corresponding SRL processes (MC.Orientation, MC.Planning, MC.Monitoring, MC.Evaluation, LC.First reading, LC.Rereading and HC.Elaboration/Organisation, detailed
definitions in the Supplemental Document). 
We defined $17$ learning actions such as RELEVANT\_READING, WRITE\_ESSAY and EDIT\_ANNOTATION in our Action Library, and mapped the trace data to these actions (as shown in Figure \ref{fig:srlStrategiesProcessesMapper}). 
For example, when learners used the essay window to write their essays, all the keyboard strokes were labelled as WRITE\_ESSAY. However, when learners opened the essay window and dwelt over their essays without typing, such actions were labelled as OPEN\_ESSAY. Further details on the action library, labelling process, and related technical issues are provided in the Supplemental Document (\href{https://drive.google.com/file/d/1B2bDqDTFRI6xWuCC0irJkWRTx-F0qkgl/view?usp=drive_link}{here}) 
Based on the definitions of the SRL processes, we mapped different actions or sequences of actions into seven SRL processes. We developed the Process Library  which contained $32$ different sequences of actions. For example, when a learner made a highlight or a note while reading the general instruction page for the task, we labelled the sequence GENERAL\_INSTRUCTION <-> EDIT\_ANNOTATION based on the action library. Further, this sequence of actions was mapped to the MC.Orientation since it indicates the learner's orientation on the task requirements. For another example, when the learner clicked on the timer, we labelled this action as TIMER, and interpreted it as MC.Monitoring since it indicates the learner's monitoring of the time left for completing the task. In cases where actions recorded in the trace data could not be mapped to any of the proposed sequences in the process library, we labelled those actions as No\_Process and did not include them into further analysis. There were \textit{$26,083$} SRL process instances in session 1 and \textit{$21,026$} in session 2. This protocol of trace-based measurement of SRL processes have previously been validated in \cite{fan2022towards,fan2022improving}. 

\subsection{Analysis}

To address \textbf{RQ1}, which aimed to investigate SRL processes and strategies within each session, we followed the approach proposed by Srivastava et al.  \cite{srivastava2022effects}. Using the sequences of SRL processes from $188$ students ($93$ from session 1 and $95$ from session 2), we applied a First-Order Markov Model (FOMM), generated with the pMineR R package, to group these sequences at the student–session level to ensure that each student had one sequence per session. The output of each FOMM was an adjacency matrix containing the transition probabilities between SRL processes. These transition matrices were then used as input for the Expectation-Maximization algorithm \cite{Ferreira2009} to detect meaningful SRL strategies. Each group or process model was subsequently represented as a graph.

To better understand the detected SRL strategies, we conducted exploratory sequence analyses, including frequency counts and descriptive statistics. We also examined transitions within and across SRL processes to gain deeper insight into how they unfolded. For each group, two models were generated: (i) a full FOMM process model and (ii) a summarised version in which SRL processes with a relative frequency below $10.0\%$ were filtered out.

To address \textbf{RQ2}, we examined the SRL strategy (groups) assigned to each student-session across the two sessions, enabling comparisons between sessions 1 and 2.
To capture changes, we visualised students’ transitions from session 1 to session 2 and applied McNemar-Bower's exact test, as the data involved repeated measures from the same students. Since the goal was to assess consistency across sessions, only students who participated in both sessions were included ($N=87$), excluding six who attended session 1 only.

To address \textbf{RQ3}, which examined the association between students' learning outcomes and their SRL strategies, we used both descriptive and inferential statistical analyses. First, we summarised descriptive statistics, including mean values and standard deviations, for each student strategy group across the learning outcomes (pre-test, post-test, and essay score). We then examined how the different SRL strategies compared with each other, using regression analyses that tested pairs of strategies (for example, S1 compared with S3). 
This analysis was conducted separately for each learning outcome (pre-test, post-test, and essay score) as the dependent variable. 
Regression coefficients (Coef) indicated the magnitude and direction of the differences, while Bonferroni-corrected P-values and effect sizes were applied to control for potential Type I errors resulting from multiple comparisons of SRL strategies \cite{Armstrong2014}. 

Incomplete observations with missing learning outcomes (pre/post-test or essay scores) were excluded, leaving $64$ students in session 1 and $68$ in session 2. Most missing data related to post-tests. To check for bias, we analysed the excluded students ($N = 56$) and found no differences in mean essay scores or prior knowledge, suggesting comparable task effort to those included.





\section{Results}


\subsection{RQ1. Self-regulated processes and strategies in secondary students' essay writing}

\begin{figure*}[htbp]
    \centering
    \includegraphics[width=1\linewidth]{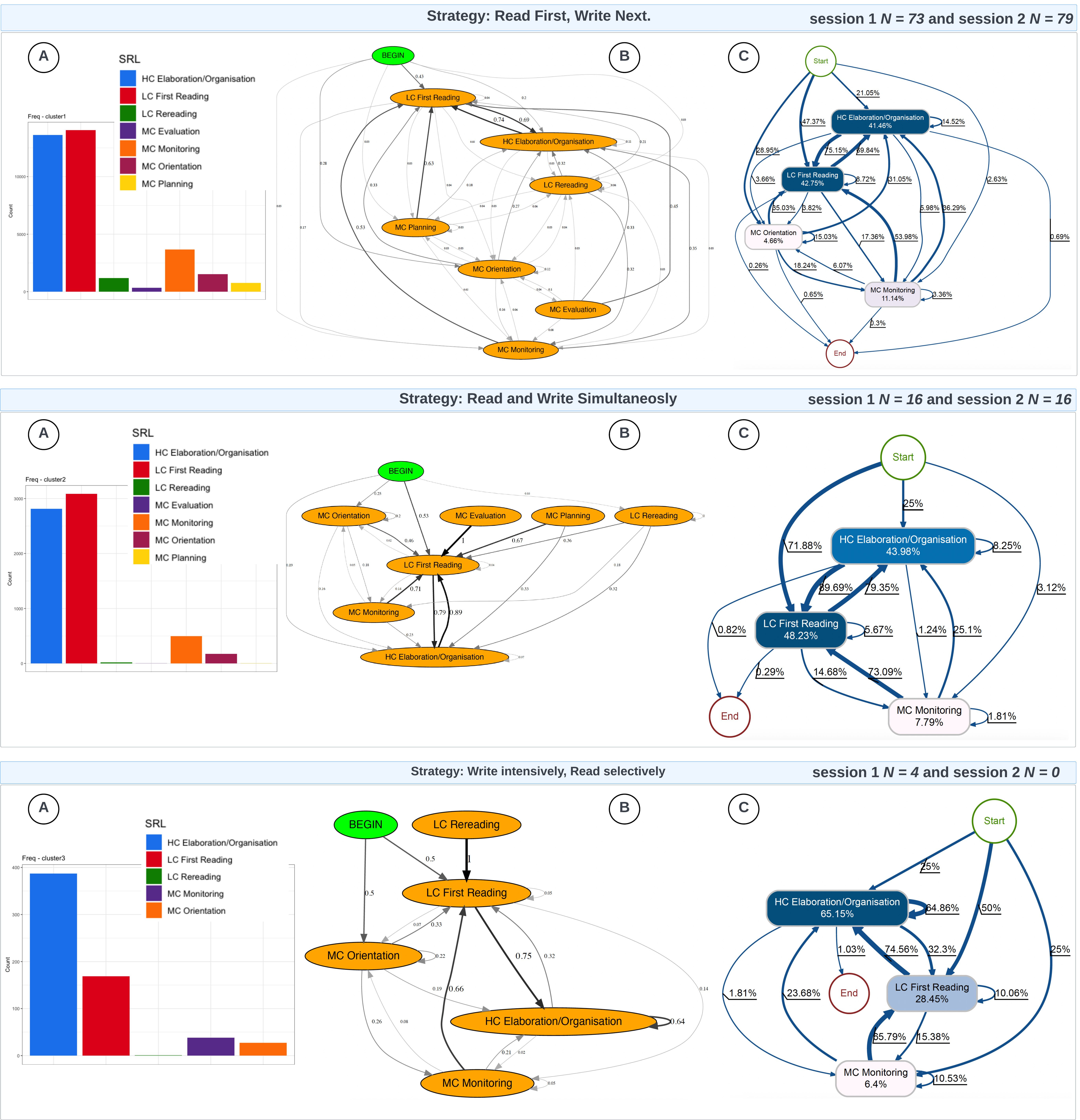}
    \caption{Three SRL strategies were identified using the First-Order Markov Model (FOMM) and Expectation-Maximisation clustering analysis. For each strategy or group: (A) frequency of SRL processes, (B) First-Order Markov Model of SRL processes, and C) Summarised First-Order Markov Model model, filtering out SRL process with a relative frequency below $10\%$.}
    \label{fig:Session1Clusters}
\end{figure*}

Session 1 had an average (mean) duration of $44$ minutes and $35$ seconds with an standard deviation (SD) of $21$ minutes and $55$ seconds, while session 2 had an average duration of $44$ minutes and $18$ seconds with an SD of $16$ minutes and  $18$ seconds. The descriptive statistics of the time spent on each SRL process is shown in Table \ref{tab:descriptiveAnalysisDistributionSRLprocesses}.

\begin{table*}[htbp]
\centering
\caption{Descriptive statistics of the time spent on each SRL process (in minutes).}
\label{tab:descriptiveAnalysisDistributionSRLprocesses}
\begin{tabularx}{0.75\textwidth}{{l}{c}{c}}
\toprule
\textbf{SRL process} & \textbf{Session 1} & \textbf{Session 2} \\
\midrule
HC.Elaboration/Organisation & $mean=24.14$, $SD=18.91$ &  $mean=20.76$, $SD=8.77$ \\
LC.First Reading & $mean=14.01$, $SD=8.044$ & $mean=3.67$, $SD=8.90$ \\
LC.Rereading & $mean=1.90$, $SD=2.08$ & $mean=1.86$,  $SD=2.38$ \\
MC.Evaluation & $mean=0.28$, $SD=0.33$ & $mean=1.21$, $SD=1.35$ \\
MC.Monitoring & $mean=1.92$, $SD=9.63$ & $mean=2.16$, $SD=4.15$ \\
MC.Orientation & $mean=2.78$, $SD=2.59$ & $mean=4.92$, $SD=10.99$ \\
MC.Planning & $mean=1.31$, $SD=2.39$ & $mean=3.48$, $SD=10.34$ \\
\bottomrule
\end{tabularx}
\end{table*}

Figure~\ref{fig:Session1Clusters} -- A shows the frequencies of SRL processes observed in both sessions. Figure~\ref{fig:Session1Clusters} -- B presents the groups identified through students' 
FOMM and Figure~\ref{fig:Session1Clusters} -- C summarised the FOMM process model. 

For the $N=188$ students in sessions 1 ($N=93$) and 2 ($N=95$), three SRL strategies were identified.
\textbf{Strategy: Read first, write next}. This strategy was adopted by the largest number of learners (session 1: $N = 73$ - $78.47\%$; session 2: $N = 79$ - $83.15\%$). Learners who used this strategy spent a roughly equal amount of time on \textit{HC.Elaboration/Organisation} and \textit{LC.First Reading} SRL processes (Figure \ref{fig:Session1Clusters} -- A, B, and C). The two prevalent SRL processes of this strategy can be clearly observed in Figure \ref{fig:Session1Clusters} -- A. However, the distribution shows that they initially focused more on reading and gradually substituted reading-oriented SRL processes with those focused on elaboration and organisation. The presence of other SRL processes was comparable to the other two strategies, except that the students who used this strategy showed less orientation (\textit{MC.Orientation}) and more monitoring (\textit{MC.Monitoring}) than the others. 


\textbf{Strategy: Read and write simultaneously}. This strategy  had the second largest user base among the participants (session 1: $N = 16$, $17.0\%$; session 2: $N = 16$, $16.84\%$). Frequency-wise, HC.Elaboration/Organisation was the dominant SRL process, followed by LC.First Reading (see Figure \ref{fig:Session1Clusters} -- A for this strategy). In other words, the students who adopted this strategy prioritised the elaboration or writing part of the task, whereas less time was devoted to reading (exploring reading materials and reading contents they had not encountered before). Figure \ref{fig:Session1Clusters} -- B and C suggest that the students who used this strategy were writing and reading at the same time; that is, going back and forth between the the essay and reading materials, to collect the required information and write it down. 

\textbf{Strategy: Write intensively, read selectively}. The proportion of learners who used this strategy was the smallest (session 1:  $N =4$ - $4.30\%$; session 2: $N =0$). The learners who used this strategy can be summarised as essay-orientated, selective-readers. For the learners who used this strategy, the learning task was perceived as an writing task which is evident by the high-proportion of writing processes (HC.Elaboration/Organisation) and comparatively low time devoted to the LC.First Reading SRL process (Figure \ref{fig:Session1Clusters} -- B). The learners who used this strategy were more likely (transition probability = $65.15\%$) to stay in the HC.Elaboration/Organisation process (Figure \ref{fig:Session1Clusters} -- C). This is also confirmed in the frequency plot (Figure \ref{fig:Session1Clusters} -- A), which shows that HC.Elaboration/Organisation and LC.First Reading were most frequent.

\subsection{RQ2. Consistency of self-regulated learning strategies in essay writing}

\begin{figure}[htb]
    \centering
    \includegraphics[width=1\linewidth]{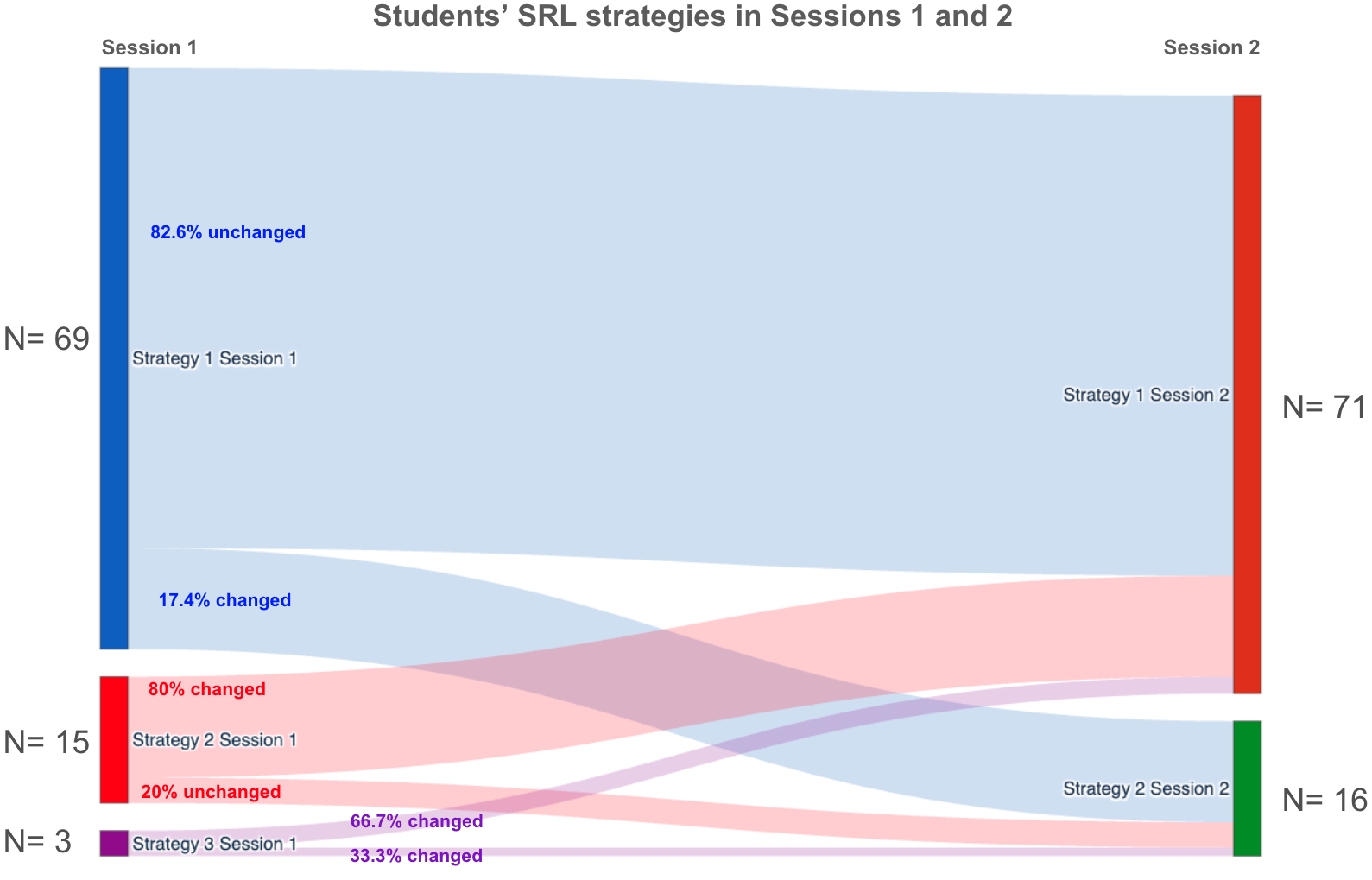}
    \caption{Sankey diagram showing students' transitions between SRL strategies from Session 1 to Session 2.}
    \label{fig:StrategyChange}
\end{figure}

Figure~\ref{fig:StrategyChange} shows shifts in SRL strategies among the $87$ students who attended both sessions; those absent from either session were excluded.

From the visualisation, most students who began with the \textit{Read first, write next} strategy ($82.61\%$) kept using it across sessions. In contrast, most who started with \textit{Read and write simultaneously} switched, mainly to \textit{Read first, write next} ($80\%$). None of the students who used \textit{Write intensively, read selectively} in session 1 continued using it in session 2; instead, they transitioned to the other two strategies, although only three students were classified to have used the \textit{Write intensively, read selectively} strategy. Figure~\ref{fig:StrategyChange} summarises these changes, with percentages showing the proportion of students shifting strategies across sessions.  


The large proportion of students ($82\%$) who remained in the \textit{Read first, write next} strategy across sessions, points to a tendency for consistency within this strategy. 
The McNemar–Bowker's symmetry test on paired SRL strategies of each student across sessions ($\chi^{2}=3.0$, $p=0.392$) suggested that changes between strategies were balanced, with no evidence of a directional preference, students were equally likely to switch from one strategy to another.

\subsection{RQ3. Impact of Strategies on Scores and Learning Outcomes} 

To examine whether strategies affected learning outcomes, Table \ref{tab:descriptivestatistics} summarises the descriptive statistics of students' results for the pre-test, post-test, and essay scores. The table aggregates results from session 1 ($64$ students with complete information on learning outcomes) and session 2 ($68$ students with complete information). Table \ref{tab:descriptivestatistics} shows that the students who used the \textit{Write intensively,
read selectively} strategy appear to have been the most effective, as students achieved highest pre-test ($0.57$), post-test ($0.82$), and essay scores ($12.33$) compared to the other strategies. The use of the \textit{Read first, write next} followed the use of the \textit{Write intensively, read selectively} strategy in terms of effectiveness, while the \textit{Read and write simultaneously} was associated with the lowest performance across learning outcome measures. 

\begin{table*}[htbp]
\centering
\scriptsize
\caption{Descriptive analysis. Pre-test and Post-test (normalised to a 0–1 scale) and Essay Scores by SRL Strategy.}
\label{tab:descriptivestatistics}
\begin{tabularx}{\textwidth}{lXXX}
\toprule
 & \textbf{Strategy 1: Read first, write next} & \textbf{Strategy 2: Read and write simultaneously} & \textbf{Strategy 3: Write intensively, read selectively} \\ 
\cmidrule(lr){2-4}
 & \textbf{Sessions 1 and 2 (N=108)} & \textbf{Sessions 1 and 2 (N=21)} & \textbf{Sessions 1 and 2 (N=3)} \\
\midrule
\textbf{Essay Score} & $mean=9.81$, $SD=2.62$ & $mean=9.19$, $SD=2.01$ & $mean=12.33$, $SD=4.72$ \\
\textbf{Pre-test score} & $mean=0.53$, $SD=0.17$ & $mean=0.46$,$ SD=0.14$ & $mean=0.578$, $SD=0.13$ \\
\textbf{Post-test score} & $mean=0.58$, $SD=0.20$ & $mean=0.498$, $SD=0.16$ & $mean=0.82$, $SD=0.10$ \\
\bottomrule
\end{tabularx}
\end{table*}

The results of the pairwise statistical comparisons of students' SRL strategies for essay scores are summarised in Table \ref{tab:statisticsScoreLG}. For essay scores, the statistical analysis confirmed the patterns observed in the descriptive analysis. In row 1 of Table \ref{tab:statisticsScoreLG}, when compared to strategy \textit{Read first, write next} (S1), students who used strategy \textit{Read and write simultaneously} (S2) scored lower ($-0.62$ points), though the difference was not statistically significant. When the use of strategy \textit{Read first, write next} (S1) was compared with strategy \textit{Write intensively,
read selectively} (S3), students who used the latter one scored higher ($2.51$ points), but the difference was not statistically significant, see row 2 in Table \ref{tab:statisticsScoreLG}. 
Finally, when the use of strategy \textit{Read and write simultaneously} (S2) was compared with the use of strategy \textit{Write intensively, read selectively} (S3), students who used  the latter one scored higher ($3.14$ points), and the difference was statistically significant ($p = 0.045$ with the large effect size), see row 3 in Table \ref{tab:statisticsScoreLG}.

\begin{table*}[htbp]
\centering
\caption{The results of the pairwise comparisons of the use of SRL strategies for essay scores and learning gains. S1:\textit{Read first, write
next}, S2:\textit{Read and write simultaneously}, and S3:\textit{Write intensively, read selectively}}
\label{tab:statisticsScoreLG}
\begin{tabularx}{\textwidth}{lXXXXXXX}
\toprule
& \multicolumn{7}{c}{\textbf{Essay Score}} \\
\cmidrule(lr){2-8}
Pair & Coef & t & P$>$|t| & Corrected P & Cohen's d & \multicolumn{2}{c}{$[0.025,~~ 0.975]$} \\
\midrule
S1 vs S2 & -0.6286 & -1.036 & 0.302 & 0.906 & 0.247 & -1.829  & 0.572 \\
S1 vs S3 &  2.5143 &  1.601 & 0.112 & 0.336  & -0.938  & -0.599 & 5.627 \\  
S2 vs S3 &  3.142 &  2.129 & \textbf{0.045} & 0.135  & -1.313  & 0.081  & 6.205 \\
\bottomrule
\end{tabularx}
\end{table*}


The results of the pairwise statistical comparisons of students' SRL strategies for pre-test and post-test are summarised in Table \ref{tab:statisticsPrePostTests}. For the pre-test, students who used the \textit{Read and write simultaneously -- S2} strategy  were more likely to perform lower ($-0.069$ points) than students who used the \textit{Read first, write next -- S1} strategy. Students who used the \textit{Write intensively, read selectively -- S3} strategy tended to perform higher ($0.042$ points) compared to students who used the \textit{Read first, write next -- S1} strategy. Finally, students who used  the \textit{Write intensively, read selectively - S3} strategy were more likely to perform better than students who used the \textit{Read and write simultaneously -- S2} strategy. However, none of these results were statistically significant. 

\begin{table*}[htbp]
\centering
\scriptsize
\caption{The results of the pairwise comparisons of the use of SRL strategies for pre-test and post-test scores. S1:\textit{Read first, write next}, S2:\textit{Read and write simultaneously}, and S3:\textit{Write intensively, read selectively}}
\label{tab:statisticsPrePostTests}
\begin{tabularx}{\textwidth}{l*{7}{c}*{7}{c}}
\toprule
& \multicolumn{7}{c}{\textbf{Pre Test}} & \multicolumn{7}{c}{\textbf{Post Test}} \\
\cmidrule(lr){2-8} \cmidrule(lr){9-15}
Pair & Coef & t & P$>$|t| & Corrected P & Cohen's d & \multicolumn{2}{c}{[0.025  0.975]} 
& Coef & t & P$>$|t| & Corrected P & Cohen's d & \multicolumn{2}{c}{[0.025~~  0.975]} \\
\midrule
S1 vs S2 & -0.0690 & -1.679 & 0.096 & 0.288 & 0.400  & -0.150 &  0.012
         & -0.0829 & -1.740 & 0.084 & 0.252 & 0.415  & -0.177 & 0.011  \\
S1 vs S3 & 0.0429  & 0.416  & 0.678 & 1 & -0.243 & -0.161   & 0.247   
         & 0.2409  & 2.010  & \textbf{0.047} & 0.141 & -1.176 & 0.003 &  0.478 \\
S2 vs S3 & 0.1119  & 1.244  & 0.226 & 0.678 & -0.767 & -0.075   &  0.298 
         & 0.3238 & 3.350  & \textbf{0.003} & 0.009 & -2.06  & 0.123  &  0.524 \\
\bottomrule
\end{tabularx}
\end{table*}

For the post-test, students who used the \textit{Read and write simultaneously -- S2} strategy performed lower ($-0.082$ points) compared to those who used the \textit{Read first, write next -- S1} strategy, although this difference was not statistically significant (see Table \ref{tab:statisticsPrePostTests}, row 1). When compared to the users of the \textit{Read first, write next -- S1} strategy, the students who used the \textit{Write intensively, read selectively -- S3} strategy performed slightly better ($+0.24$ points) (see Table \ref{tab:statisticsPrePostTests}, row 2), and it was statistically significant ($p=0.047$) with a large effect size (d=-1.175). Finally, when compared to the users of the \textit{Read and write simultaneously -- S2} strategy, students who used the \textit{Write intensively, read selectively -- S3} strategy also performed better ($+0.32$ points) (see Table \ref{tab:statisticsPrePostTests}, row 3), and it was also statistically significant ($p=0.003$) with a large effect size (d=-2.06). The pre- and post-test results in this analysis confirmed the descriptive findings: the use of the \textit{Write intensively, read selectively -- S3} strategy appears to have been the most effective for these learning outcomes, followed by the \textit{Read first, write next -- S1} strategy, and then the \textit{Read and write simultaneously -- S2} strategy.

\section{Discussion}

\subsection{Repeated Measures of Secondary School Students’ SRL Strategies}

For \textbf{RQ1}, we found that students tended to demonstrate three SRL strategies: (1) \textit{Read first, write next}, (2) \textit{Read and write simultaneously}, and (3) \textit{Write intensively, read selectively}. These strategies were observed in both sessions, with most students following the \textit{Read first, write next} strategy (in session 1, $N = 73$, and in session 2, $N = 79$), followed by the \textit{Read and write simultaneously} strategy (in session 1, $N = 16$, and in session 2, $N = 16$), and finally the \textit{Write intensively, read selectively} strategy (in session 1,  $N =4$, and in session 2, $N =0$). These findings are consistent with the work of Srivastava et al.  \cite{srivastava2022effects}, who also identified these three SRL strategies for higher education contexts. 
Although Srivastava et al. \cite{srivastava2022effects} conducted their study with higher education students, the distribution of students across strategies is similar to what we observed. Our findings also partially align with Lämsä et al. \cite{Lamsa2025}, who identified four strategies in secondary education also for multi-source essay writing: \textit{Writing with metacognitive monitoring}, \textit{Writing intensively}, \textit{Reading first, writing next}, and \textit{Reading and writing simultaneously}. Unlike the \cite{Lamsa2025} study, we did not observe a strategy characterised by intensive or moderate use of metacognitive monitoring. Nevertheless, monitoring was present, though not prevalent, across all strategies we identified. The differences between our results and those of Lämsä et al. \cite{Lamsa2025} may stem from methodological differences. Whereas their study relied on automated detection of SRL strategies from learning actions derived from raw log data, our approach drew on theorised SRL processes identified through a process library, which allowed us to map sequences of learning actions to higher-order constructs in a more conceptually grounded way. These findings highlight opportunities for further research to compare clustering approaches and examine how students’ monitoring behaviours influence the definition and weighting of strategies grounded in SRL processes.

Our analysis of secondary school students’ SRL strategies revealed particularly limited engagement in metacognitive processes, especially monitoring and control. Alnashiri et al. \cite{Halima2025} also found that secondary school students demonstrated few metacognitive processes (e.g., monitoring and control), and our results extend these findings by providing novel insights on the SRL strategy level. Across the three SRL strategies, students showed limited instances of monitoring. For example, in Figure~\ref{fig:Session1Clusters}--C students who used the \textit{Read first, write next} strategy engaged in monitoring $11.14\%$ of the time, users of \textit{Read and write simultaneously} only $7.79\%$, and users of \textit{Write intensively, read selectively} $6.4\%$. 
Overall, these results suggest that students in secondary schools lack strategies for planning, monitoring, and evaluating their writing independently. This concern is amplified in the context of GenAI, where learners may offload substantial parts of their reading and writing work to the AI system, leading to reduced metacognitive engagement — a phenomenon often described as ‘metacognitive laziness’ \cite{fan2025beware}. This highlights the need to design support mechanisms that can help students strengthen these metacognitive processes and become more self-regulated during their writing tasks. 

Although our findings showed limited monitoring and control, earlier research found that secondary students often engaged in Orientation, Re-reading, and Elaboration/Organisation  Cheng et al.  \cite{cheng2025self}. Our analysis revealed that Elaboration/Organisation was prevalent across all three strategies, while Orientation and Re-reading, though present (Figure~\ref{fig:Session1Clusters}--B), were not frequently enacted. These discrepancies underscore the need for future comparative studies that can capture how SRL processes manifest differently across strategies and contexts, offering a deeper understanding of the dynamic nature of SRL.

\subsection{Consistency of Secondary School Students’ SRL Strategies Across Repeated Measures}

For \textbf{RQ2}, our results showed variability in strategy use: many students remained using  the \textit{Read first, write next} strategy ($82.60\%$), most students shifted from the \textit{Read and write simultaneously} strategy to the \textit{Read first, write next} strategy ($80\%$), and all students moved from the \textit{Write intensively, read selectively} strategy to other strategies in session 2. 
These results align with the findings of Matcha et al. \cite{matcha2019analytics}, who reported that students' strategies tend to diversify over the course of a semester. However, further research is needed to explore how the approach used in our study can be extended to other learning tasks and contexts (e.g., entire school year), given the dynamic \cite{matcha2019analytics} and context-dependent nature of SRL strategies \cite{uzir2020analytics, fan2021learning, van2010development,rakovic2023network}.

\subsection{Secondary School Students’ SRL Strategies and Learning Outcomes}

For \textbf{RQ3}, we found an association between learning outcomes and the SRL strategies adopted by students. In particular, those who used the \textit{Write intensively, read selectively} strategy achieved higher essay scores. As shown in Table~\ref{tab:statisticsScoreLG}, pairwise comparisons revealed a statistically significant difference between this strategy and the \textit{Read and write simultaneously} strategy ($p=0.045$). Moreover, students who adopted the \textit{Write intensively, read selectively} strategy showed greater gains in prior knowledge compared to the other two strategies, with results reaching statistical significance (Table~\ref{tab:statisticsPrePostTests}).    

Our findings suggest that the \textit{Write intensively, read selectively} strategy was associated with higher essay scores, whereas \citet{srivastava2022effects} reported poorer outcomes for the same strategy. A possible explanation for this discrepancy is that students who followed the \textit{Write intensively, read selectively} strategy in our study exhibited a more balanced and diverse selection of SRL processes (e.g., Fig. \ref{fig:Session1Clusters}--C shows Reading, Orientation, Elaboration, and Monitoring), a characteristic that has previously been linked to better learning outcomes~\cite{bannert2014process}. Therefore, while the use of \textit{Read first, write next} can be effective when students understand the task and engage in continuous monitoring, in our study this strategy (Fig. \ref{fig:Session1Clusters}--C) was more often associated with lower learning outcomes compared \textit{Write intensively, read selectively}.


Although re-reading was present in all strategies, it was more frequent in \textit{Read first, write next}. Yet these frequencies did not provide enough evidence to conclude that re-reading directly impacts learning outcomes. Prior work found re-reading common among high-performing secondary students~\cite{cheng2025self}, but also showed that low-performing students often used strategies with frequent transitions between Organisation/Elaboration and Monitoring, signalling poor task understanding. In our study, the \textit{Write intensively, read selectively} strategy involved frequent orientation and consistent monitoring. This combination may indicate that students were deliberately selecting effective strategies to achieve high performance. These findings highlight a promising direction for future research on the role of orientation and monitoring in SRL strategies and their impact on learning outcomes.

\subsection{Implications and limitations}

For Learning Analytics, our findings demonstrate that applying a First-Order Markov Model in combination with the Expectation-Maximization algorithm \cite{Ferreira2009} to sequences of SRL processes is a promising approach for consistently identifying SRL strategies in secondary schools. This methodological contribution extends prior work in higher education \cite{srivastava2022effects} and shows that process-level analyses can be scaled to younger learners. Such approaches can inform the design of analytics-driven feedback tools and dashboards that move beyond simple activity counts to provide strategy-level insights, thereby enabling more actionable support for students and teachers. In particular, the results highlight the need for LA systems to better capture and scaffold metacognitive processes—such as planning, monitoring, and evaluation—that were found to be limited in secondary school students’ strategies.

The  evidence  reported  in this paper  should  be  considered  in  the context  of  the  limitations  of  the  study. A limitation of our study is the lack of insight into students' internal conditions that shaped their choices for enacting different strategies across sessions. Future studies should incorporate measures of students' internal conditions (e.g., metacognitive knowledge) for engaging with specific learning tasks. An alternative approach is illustrated in the work of Srivastava et al. \cite{srivastava2022effects}, who examined both internal factors (e.g., student motivation) and external conditions (e.g., interventions through feedback or scaffolds). Second, the data were collected from a short, two-session task, which may not fully represent how strategies evolve over longer timeframes or across different learning contexts. Finally, although our clustering approach revealed meaningful strategy groups, comparisons with alternative approaches (e.g., automated detection from raw learning actions) are needed to validate the robustness of these findings. Future research should integrate additional data sources, including self-reports, motivation measures, or classroom interventions, to provide a more comprehensive account of why students shift strategies and how learning analytic tools can support more productive forms of self-regulated learning.

\section{Conclusion}

This study examined how secondary school students enact self-regulated learning (SRL) strategies during online essay writing, how consistently these strategies are applied across repeated tasks, and how they relate to learning outcomes. Using trace data collected across two writing sessions and applying process mining and unsupervised machine learning techniques, we identified three dominant SRL strategies grounded in established SRL processes: \textit{Read first, write next}, \textit{Read and write simultaneously}, and \textit{Write intensively, read selectively}.

The findings show that most students consistently relied on the \textit{Read first, write next} strategy across sessions, indicating stability in their approach to writing from multiple sources. However, strategy effectiveness varied. Although less frequently adopted and not sustained across sessions, the \textit{Write intensively, read selectively} strategy was most strongly associated with essay performance and post-test improvement. This suggests that learning success is shaped not by how often strategies are used, but by how SRL processes are coordinated during learning.

Across all strategies, students demonstrated limited engagement in metacognitive processes such as planning, monitoring, and evaluation. This underscores a critical challenge in secondary education, where learners are still developing the metacognitive skills required for effective self-regulation. In increasingly digital learning environments, and with the growing presence of generative AI tools, these findings highlight the need for scaffolds that strengthen metacognitive regulation rather than simply supporting task completion.

Overall, this study extends prior SRL research by providing rare repeated-measures evidence from secondary education and demonstrating the value of strategy-level analyses for understanding how students learn online. Future research should explore how SRL strategies evolve across varied tasks, longer timeframes, and instructional conditions, and how analytics-informed scaffolds can support the development of productive self-regulated learning during adolescence.

\bibliographystyle{unsrt}  
\bibliography{bibliography}  

\end{document}